\documentclass{iaus}
\usepackage{graphicx}
\usepackage{natbib}
\usepackage{epsfig}

\newcommand{\mH}{\rm{H}}
\newcommand{\Hm}{\rm{H^{-}}}
\newcommand{\me}{\rm{e^{-}}}
\newcommand{\Hp}{\rm{H^{+}}}
\newcommand{\He}{\rm{He}}
\newcommand{\mHt}{\rm{H}_{2}}
\newcommand{\hd}{\rm{HD}}
\newcommand{\lih}{\rm{LiH}}
\newcommand{\li}{\rm{Li}}
\newcommand{\htp}{\rm{H}_{3}^{+}}
\newcommand{\mHtp}{\rm{H}_{2}^{+}}

\def\simless{\mathbin{\lower 3pt\hbox
   {$\rlap{\raise 5pt\hbox{$\char'074$}}\mathchar"7218$}}}
\def\simgreat{\mathbin{\lower 3pt\hbox
   {$\rlap{\raise 5pt\hbox{$\char'076$}}\mathchar"7218$}}}

\title[Open questions in Pop.~III star formation] 
{Open questions in the study of population~III star formation}

\author[Glover \etal]   
{S.~C.~O.~Glover$^{1,2}$, P.~C.~Clark$^2$, T.~H.~Greif$^2$, J.~L.~Johnson$^3$,
\\ V.~Bromm$^3$, R.~S.~Klessen$^2$, \& A.~Stacy$^3$}

\affiliation{$^1$Astrophysikalisches Institut Potsdam, An der Sternwarte 16, 14482 Potsdam, Germany \\
[\affilskip]
$^2$ Institut f\"{u}r theoretische Astrophysik, Albert-Ueberle Strasse 2, 69120 Heidelberg, Germany \\
[\affilskip]
$^3$ Department of Astronomy, University of Texas, Austin, TX 78712 \\}

\pubyear{2008}
\volume{255}  
\pagerange{1}
\jname{Low-Metallicity Star Formation: From the First Stars to Dwarf Galaxies}
\editors{L.K. Hunt, S. Madden \& R. Schneider, eds.}
\begin{document}

\maketitle

\begin{abstract}
The first stars were key drivers of early cosmic evolution.
We review the main physical elements of the current consensus
view, positing that the first stars were predominantly very
massive. We continue with a discussion of important open
questions that confront the standard model. Among them are 
uncertainties in the atomic and molecular physics of the
hydrogen and helium gas, the multiplicity of stars that form in 
minihalos, and the possible existence of two separate modes of 
metal-free star formation.

\end{abstract}

\firstsection 
\section{Introduction}
We have learnt a great deal over the last decade concerning the formation of the first
stars, the so-called population III or Pop.~III. A consensus has emerged regarding
the properties of the protogalaxies (or `minihalos')
in which the first stars formed and the major physical
and chemical processes involved in their formation. Their masses remain uncertain, but
are widely expected to be significantly larger than the characteristic mass for present-day
star formation. Good summaries of the present state of the field can be found in 
\citet{blar04}, \citet{glo05} and \citet{nm08}. Nevertheless, there remain some important open 
questions. In this review, we discuss four of the most important of these issues and the 
efforts being made to resolve them. These issues are the impact of chemical rate coefficient
uncertainties on the accuracy of our models of Pop.~III star formation; the perennial question
of whether we have identified all of the important physical processes responsible for cooling
the gas; the number of population III stars that form in each minihalo; and the question
of whether population III actually consists of two sub-populations (Pop.~III.1 and Pop.~III.2)
with different characteristic masses. Two further important issues -- the question of what
physical process terminates accretion onto the earliest protostars, and the impact of dark
matter decay and annihilation on the formation of the first stars -- are not discussed here 
as they are covered in detail elsewhere in these proceedings (see e.g.\ the contributions by
Whalen, Tan, Freese \& Iocco).

\section{Population III star formation: the consensus view}
In the $\Lambda$CDM model for cosmic structure formation, the first gravitationally
bound structures to form are very small (e.g.\ $M \sim M_{\oplus}$ if cold dark matter consists
of neutralinos; \citealt{dms05}) and consist purely of dark matter. Larger bound structures
form through accretion and through the merger of these smaller objects, in a process
known as hierarchical clustering. Once the mass of these gravitationally bound dark 
matter objects (conventionally referred to as dark matter halos) exceeds the 
cosmological Jeans mass, pressure forces can no longer prevent gas from falling into
the potential wells created by the dark matter. Infalling gas is heated by adiabatic 
compression and by shocks, and in the absence of radiative cooling it will eventually 
reach a state of hydrostatic equilibrium, with a mean temperature given by the virial
temperature of the halo, $T_{\rm vir}$. The first dark matter halos to have masses
greater than the cosmological Jeans mass have virial temperatures much smaller
than the $\sim 10^{4} \: {\rm K}$ temperature at which cooling from electronic excitation 
of atomic hydrogen becomes effective, and so the gas in these halos must rely on other, 
less effective forms of cooling. 

It has long been recognized that the most important coolant at $T < 10^{4} \: {\rm K}$
in primordial gas is molecular hydrogen, $\mHt$ \citep{sz67,pd68}. In the absence of dust,
this forms in the gas phase via the reaction chains
\begin{eqnarray}
\mH + \me & \rightarrow & \Hm + \gamma, \\
\Hm + \mH & \rightarrow & \mHt + \me,  \label{ad}
\end{eqnarray}
and
\begin{eqnarray}
\mH + \Hp & \rightarrow & \mHtp + \gamma, \\
\mHtp + \mH & \rightarrow & \mHt + \Hp,
\end{eqnarray}
with the $\Hm$ mechanism generally dominating. Gas-phase $\mHt$ formation 
therefore relies on the presence of free electrons and protons, and the amount
of $\mHt$ that can be formed by these reactions is limited by the recombination 
of the gas. A number of studies have examined the conditions required in 
order for the gas to form enough $\mHt$ to be able to cool within a Hubble time
\citep[see e.g.][]{htl96,teg97,yosh03}, with the consensus being that 
virial temperatures $T_{\rm vir} \simgreat 1000 \: {\rm K}$ are required,
corresponding to halo masses of a few times $10^{5} \: {\rm M_{\odot}}$ or 
more.

The redshift at which the first halos of the required mass are formed depends to
some extent of the definition of `first'. The very first objects of this mass to form within 
a Hubble volume do so at redshifts $z \sim 50$--60 \citep{reed05,nnb06}, but are 
exceedingly rare; on the other hand, the first object to form within a reasonable local 
volume, say 1 comoving Mpc$^{3}$, does so at a somewhat lower redshift $z \sim 30$
\citep{glo05}. For technical reasons, numerical simulations of population III star formation 
typically focus on the latter case.

If enough $\mHt$ can be formed to efficiently cool the gas, then it will undergo
gravitational collapse. Initially, this collapse occurs rapidly. The gas temperature
drops as the density increases, and so the gas becomes increasingly gravitationally
unstable. However, as the collapse proceeds, cooling from $\mHt$ becomes steadily 
less efficient. The exponential fall-off in the $\mHt$
cooling rate at low temperatures prevents it from cooling the gas below 
$T \sim 200 \: {\rm K}$,  while the approach of the rotational and vibrational level
populations of  $\mHt$ to their local thermodynamic equilibrium (LTE) values at 
densities above $n_{\rm cr} \sim 10^{4} \: {\rm cm^{-3}}$ renders $\mHt$ cooling
inefficient at higher densities. The gas therefore accumulates in a quasi-hydrostatic
``loitering state'' with a characteristic temperature $T_{\rm char} = 200 \: {\rm K}$
and characteristic density $n_{\rm char} = 10^{4} \: {\rm cm^{-3}}$
\citep{abn02,bcl02}. Eventually,
the amount of mass accumulated at this density exceeds the Bonnor-Ebert mass,
which at this point is a few hundred M$_{\odot}$, 
and the gravitational collapse resumes. However, beyond this point
$\mHt$ cooling is unable to maintain the temperature at $T = 200 \: {\rm K}$; it
steadily reheats, evolving with an effective polytropic index of 
$\gamma_{\rm eff} = 1.1$ \citep{on98}.

At densities $n > 10^{8} \: {\rm cm^{-3}}$, three-body formation of $\mHt$ via the
reactions 
\begin{eqnarray}
\mH + \mH + \mH & \rightarrow & \mHt + \mH, \label{3bh1} \\
\mH + \mH + \mHt & \rightarrow & \mHt + \mHt, \label{3bh2}
\end{eqnarray}
becomes effective and rapidly converts all of the hydrogen to molecular form 
\citep{pss83}. Although the dramatic increase in the $\mHt$ abundance boosts
the $\mHt$ cooling rate, the high gas density, the growing optical depth in the
$\mHt$ rotational and vibrational lines \citep[see e.g.][]{on98,ripa02,ra04} and the heat input 
from $\mHt$ formation all combine to prevent the temperature from dropping significantly.

At densities $n \sim 10^{14} \: {\rm cm^{-3}}$ and above, a second form of $\mHt$ cooling
becomes important: $\mHt$ collision-induced emission (CIE). Although $\mHt$
has no permanent dipole moment, the complexes formed in collisions of $\mHt$
with $\mH$, $\mHt$ or $\He$ can act as `super-molecules', with non-zero dipole 
moments. Although these excited complexes last for only a very short time
($t_{\rm coll} \sim 10^{-12} \: {\rm s}$; \citealt{ra04}), there is
nevertheless always a small probability that a photon will be emitted.
At very high densities, collisions occur frequently enough to make this emission 
a viable means of cooling the gas. However, this CIE-dominated phase lasts for
only a short time before the gas becomes optically thick in the continuum, and so
CIE cooling is unable to significantly reduce the temperature.

At even higher densities ($n > 10^{16} \: {\rm cm^{-3}}$), the gas becomes increasingly
optically thick, and so radiative cooling becomes completely ineffective. However, 
collisional dissociation of $\mHt$ acts as a heat sink, keeping the temperature evolution
of the gas close to isothermal until most of the $\mHt$ has been destroyed \citep{om05,
tao08}. This occurs
by the time that the density reaches $10^{21} \: {\rm cm^{-3}}$, and beyond this point the 
evolution of the gas becomes adiabatic. This is the moment at which we first have something 
that we can identify as a true protostar.

At the moment that this protostar forms, its mass is less than $0.01 \: {\rm M_{\odot}}$, but it is 
surrounded by a dense, massive envelope of infalling gas with a mass of hundreds of solar 
masses. Accretion of this envelope is expected to occur at a rapid rate: a simple scaling 
argument suggests that the accretion rate should scale as $\dot{M} \propto c_{\rm s}^{3} 
\propto T^{3/2}$, and since the temperature of the gas is between 10--100 times larger than in 
local star forming regions, the expected accretion rates are orders of magnitude larger than the 
rate of order $10^{-5} \: {\rm M_{\odot}} \: {\rm yr^{-1}}$ that is typical locally. Accretion rates
have been estimated using a variety of techniques \citep[see][section 4]{glo05}, and although
the estimated rates differ somewhat, in every case one expects the star to be able to accrete more 
than $100 \: {\rm M_{\odot}}$ of gas within the Kelvin-Helmholtz relaxation time. Moreover, 
stellar feedback in the form of radiation-driven winds is expected to be far less effective in
primordial gas than in metal-enriched gas \citep[see e.g.][]{kud02}, and so there seems little to 
prevent the star from becoming massive.
 
Hence, population III stars are expected to be massive and short-lived, surviving for only a 
few million years. How they end their lives depends on both their mass and the speed at which
they are rotating. Non-rotating population III stars with masses in the range
10--$50 \: {\rm M_{\odot}}$ are expected to explode as either conventional type II supernovae or
as hypernovae \citep{un02}. For masses between 50 and $140 \: {\rm M_{\odot}}$ and above $260 \:
{\rm M_{\odot}}$, direct collapse to form a black hole is expected, while non-rotating Pop.~III
stars with masses in the range 140--260$\: {\rm M_{\odot}}$ are expected to end their lives as
pair-instability supernovae \citep{hw02}. Rapid rotation changes this picture somewhat, by enhancing
the effects of mass-loss, and by inducing mixing within the star. The latter effect leads to the
star having a larger helium core at the end of its main sequence evolution, and so likely allows
less massive stars to become pair instability supernovae \citep{eks08}. 

\section{Open questions}

\subsection{Are uncertainties in the chemical rate coefficients important?}
Our ability to construct accurate models of the chemical evolution of metal-free gas 
is constrained by the level of accuracy with which the rate coefficients of the key 
chemical reactions have been determined. This varies significantly depending on the
reaction in question. Many of the most important reactions involved in the formation
and destruction of $\mHt$ have rate coefficients that have been determined to within
an accuracy of the order of 10--20~\% at typical protogalactic temperatures 
\citep{aanz97,gp98}. However, there are several important reactions whose rates 
are far more uncertain.

One example is the charge transfer reaction 
\begin{equation}
\mHt + \Hp \rightarrow \mHtp + \mH,
\end{equation}
which is a major destruction pathway for $\mHt$ in hot, ionized gas. \citet{sav04}
present a new calculation of the rate of this reaction, and show that previous
determinations, some of which remain widely used in the literature, differ by 
orders of magnitude. Fortunately, this process is unimportant  in cold gas, owing
to its large endothermicity, and so the large uncertainty in this reaction has little 
impact on the accuracy with which we can model the formation of the first stars.
However, its impact on the formation of so-called Pop.~III.2 stars (discussed in
more detail in \S\ref{pop32} below) may be much larger and deserves further study. 

Another source of uncertainty stems from the competition between two of the main
destruction pathways for $\Hm$, associative detachment (reaction~\ref{ad} above)
and mutual neutralization
\begin{equation}
\Hm + \Hp \rightarrow \mH + \mH.
\end{equation}
\citet{gsj06} surveyed the literature available on the rates of these reactions at low
temperatures ($T < 10^{4} \: {\rm K}$), and showed that both were uncertain by an
order of magnitude. They also examined the effects of this uncertainty on the chemistry,
cooling and dynamics of the gas. In the conventional Pop.~III formation scenario, the 
fractional ionization of the gas is small enough to ensure that associative detachment 
always  dominates over mutual neutralization, and so any uncertainty in the rate 
coefficients has almost no effect. In gas cooling from a highly ionized state, however,
as in some of the Pop.~III.2 scenarios discussed in section~\ref{pop32}
below, mutual neutralization
dominates initially, with associative detachment becoming important only once the gas
has recombined sufficiently. In this case, any uncertainty in the rates of these two reactions
leads to an uncertainty in the $\mHt$ formation rate, and in the final amount of $\mHt$
formed. More recently, \citet{ga08} have shown that this uncertainty affects the amount 
of HD that can form, and so also influences the minimum temperature that the collapsing
gas can reach. 

Fortunately, experimentalists have begun to address this source of uncertainty. Recent
measurements of the mutual neutralization reaction rate at low temperatures by Xavier
Urbain have reduced what was an order of magnitude uncertainty to something closer
to a 50\% uncertainty (X.\ Urbain, private communication). At the same time, an experiment
designed to accurately measure the rate coefficient for the associative detachment reaction
has been funded and is currently under construction (\citealt{bruhns08}; D.~Savin, private communication), 
and so this source of uncertainty may also have been removed in a 
few years time.

Finally, and perhaps most importantly, a very large uncertainty exists in the rate of the 
three-body $\mHt$ formation reaction. The current state of the literature regarding the
rate of reaction~\ref{3bh1} was surveyed in \citet{glo08}, who showed that in the temperature 
range $200 < T < 2000 \: {\rm K}$  relevant for population III star formation, there is an 
uncertainty in the rate coefficient of two to three orders of magnitude. Moreover, there is 
no sign that this uncertainty is reducing: indeed, the two most recent determinations 
of the rate coefficient (by \citealt{abn02} and by \citealt{fh07}) are the two with the 
greatest disagreement. The uncertainty in the rate of reaction~\ref{3bh2} is harder to
quantify, as there have been fewer studies made of this reaction. However, if we assume
\citep[following][]{jgc67} that the rate of this reaction is 1/8th that of reaction~\ref{3bh1},
then the obvious implication is that this rate has a similarly large uncertainty.

The effects of the uncertainty in the three-body $\mHt$ formation rate on the thermal
evolution of the gas have been examined by \citet{ga08} and \citet{gs08} using highly
simplified one-zone models. These studies find that the rate coefficient uncertainties
lead to an uncertainty of approximately 50\% in the temperature evolution of the gas in 
the density range $10^{8} < n < 10^{13} \: {\rm cm^{-3}}$. The effect of this uncertainty
on the dynamical evolution of the gas and in particular on the predicted protostellar 
accretion rates are not currently known, although work is currently under way to address
this.

\subsection{Are we including all of the significant coolants?}
As previously noted, $\mHt$ has long been recognized as the most important coolant
in primordial gas at temperatures $T < 10^{4} \: {\rm K}$. At the same time, it is clear 
from our previous discussion that $\mHt$ becomes increasingly ineffective as a coolant
as we move to higher densities, owing to the low critical density at which its rotational
and vibrational level populations reach their LTE values, and to the fact that at densities
$n \simgreat 10^{10} \: {\rm cm^{-3}}$, optical depth effects further suppress $\mHt$ 
cooling. 

The comparative ineffectiveness of $\mHt$ as a coolant in high density metal-free
gas has motivated various authors to examine the role that might be played by
other coolants at high densities. Perhaps the best studied alternative coolant is hydrogen 
deuteride, HD. It has excited rotational and vibrational levels that have radiative lifetimes 
that are about a factor of 100 shorter than those of $\mHt$,  and so the $\hd$ cooling rate 
does not reach its LTE limit until $n \sim 10^{6} \: {\rm cm^{-3}}$. It is also a far more effective
coolant than $\mHt$ at low temperatures ($T \simless 200 \: {\rm K}$; see e.g., \citealt{flpr00}). 
This is due primarily to the fact that radiative transitions can occur between rotational levels 
with odd and even values of $J$, allowing cooling to occur through the $J=1 \rightarrow 0$ 
transition. The corresponding odd $\leftrightarrow$ even transitions in the case of $\mHt$ 
represent conversions from ortho-$\mHt$ to para-$\mHt$ or vice versa, and are highly 
forbidden. Furthermore, at low temperatures the ratio of $\hd$ to $\mHt$ can be significantly 
enhanced with respect to the cosmological D:H ratio by chemical fractionation \citep[see e.g.,][]{glo08}.

The role of $\hd$ cooling in early minihalos has been investigated by a number of 
authors. In the case of the earliest generation of minihalos, which form from very 
cold neutral gas that is never heated to more than a few thousand K during the course 
of the galaxy formation process, the importance of $\hd$ appears to be a function of
the size and dynamical history of the minihalo \citep{ripa07,mb08}. In small minihalos
($M \simless 10^{6} \: {\rm M_{\odot}}$) that collapse in an unperturbed, relatively uniform
fashion, $\mHt$ cooling can lower the temperature of the gas enough to allow $\hd$ 
(which is strongly enhanced at low temperatures by chemical fractionation; \citealt{glo08})
to take over and dominate the cooling. In larger minihalos ($M \simgreat 10^{6} \: {\rm M_{\odot}}$)
that have a more complex dynamical history, $\mHt$ cooling is unable to lower the temperature
to the same extent, and so the gas never becomes cold enough for HD to dominate. In
this case, it contributes no more than about 20--30\% of the total cooling \citep{gs08} and does
not appear to significantly affect the dynamics of the gas \citep{bcl02}. HD cooling 
is also of great importance in situations where an increase in the fractional ionization of the 
gas has led to an increase in the $\mHt$ fraction. In this situation, the gas often becomes
cool enough for HD to dominate \citep[see e.g.,][]{nu02,no05,jb06,sv06}. This scenario is discussed 
in more detail in \S\ref{pop32} below.

Another molecule to have attracted considerable attention is lithium hydride, LiH.
This molecule has a very large dipole moment, $\mu = 5.888 \: {\rm debyes}$
\citep{zs80}, and consequently its excited levels have very short radiative lifetimes.
Therefore, despite the very low lithium abundance in primordial gas ($x_{\rm Li} = 4.3
\times 10^{-10}$, by number; see \citealt{cy04}), it was thought for a time  that
LiH would dominate the cooling at very high densities \citep[see e.g.,][]{ls84}.
However, accurate quantal calculations of the rate of formation of $\lih$ by
radiative association \citep{dks96,ggg96,bdlg03}
\begin{equation}
 \li + \mH \rightarrow \lih + \gamma,
\end{equation}
have shown that the rate is much smaller than was initially assumed,
while recent work by \citet{dpgg05} has shown that the reaction
\begin{equation}
\lih + \mH \rightarrow \li + \mHt,
\end{equation}
has no activation energy and so will be an efficient destruction mechanism for
$\lih$ for as long as some atomic hydrogen remains in the gas. Consequently,
the amount of lithium hydride present in the gas is predicted to be very small,
even at very high densities, and so $\lih$ cooling is no longer believed to be
important \citep{mon05,gs08}.

Finally, molecular ions such as  $\mHtp$, $\htp$ or ${\rm HeH^{+}}$ provide another
possible source of cooling in dense primordial gas. Early work on $\mHtp$ 
cooling in ionized primordial gas can be found in \citet{ss77,ss78}, and its possible 
importance in hot, highly ionized conditions has recently been re-emphasized by 
\citet{yokh07}. However, it has a low critical density ($n_{\rm cr} \sim 10^{3} 
\: {\rm cm^{-3}}$) and is also readily destroyed in collisions with atomic hydrogen
\begin{equation}
\mHtp + \mH \rightarrow \mHt + \Hp.
\end{equation}
These factors make it unlikely to be an important coolant at high densities. 

${\rm HeH^{+}}$ is a more promising candidate: it has a large dipole moment,
a large cooling rate per molecule when in LTE, and hence a very large critical
density \citep{engel05}. However, once again it is readily destroyed in collisions with
atomic hydrogen
\begin{equation}
{\rm HeH^{+}} + \mH \rightarrow \mHtp + \He,
\end{equation}
and so its abundance in high density gas is very small \citep{gs08}. It therefore
never becomes a significant coolant.

The last of these three molecular ions, $\htp$, is perhaps the most interesting.
It has a large cooling rate per molecule when in LTE \citep{nmt96} and hence
a large critical density \citep{gs06,gs08}. Unlike $\mHtp$ and ${\rm HeH^{+}}$, it is
not readily destroyed by collisions with atomic hydrogen -- the reaction
\begin{equation}
\htp + \mH \rightarrow \mHtp + \mHt
\end{equation}
does occur, but must overcome a large energy barrier, and so proceeds slowly at
temperatures $T < 1000 \: {\rm K}$. Moreover, $\htp$ is known to be an important coolant 
in at least one astrophysical scenario, namely in the upper atmospheres of gas giants 
\citep{miller00}. \citet{gs08} have examined in detail the role that $\htp$ plays in
the cooling of primordial gas. They find that in most variations of the conventional
Pop.~III.1 formation scenario, $\htp$ comes close to being an important coolant, but 
never quite succeeds. It contributes to the total cooling rate at densities $10^{7}
< n < 10^{9} \: {\rm cm^{-3}}$ at
the level of a few percent, making it the third most important coolant after $\mHt$
and HD, but unimportant for the overall thermal evolution of the gas. However, 
\citet{gs08} do identify one scenario in which $\htp$ can become the dominant coolant.
If a significant ionization rate can be maintained at densities $n > 10^{8} \: 
{\rm cm^{-3}}$, for instance by cosmic rays or very hard X-rays, then ionization of
$\mHt$ to $\mHtp$ is quickly followed by the reaction
\begin{equation}
\mHtp + \mHt \rightarrow \htp + \mH
\end{equation}
resulting in the production of a large number of $\htp$ ions. In this scenario, the
$\htp$ abundance can be maintained at a high enough level to allow $\htp$ to dominate
the cooling rate. The required ionization rate is of the order of $10^{-17} \: {\rm s^{-1}}$. 
This is far 
larger than could be produced by plausible extragalactic sources, but is perhaps
consistent with production by local sources. An interesting possibility in this 
context is that dark matter annihilation within the dense core may provide the 
necessary source of ionization.

\subsection{How many stars form per minihalo?}
High resolution AMR and SPH simulations of the formation of the first stars typically find 
that only a single collapsing protostellar core forms in each minihalo
\citep[see e.g.][]{abn02,yoha06,on07}. However,
it is possible that this result is a consequence of the numerical methods used to simulate
the gas, rather than of the gas physics. In order to properly resolve the gravitational collapse 
of the gas, it is necessary to ensure that the gravitational Jeans length is resolved with sufficient
computational elements, a criterion that has been formalized by \citet{tru97} for grid-based codes and
by \citet{bb97} for SPH. If the gas remains close to isothermal during the collapse, then the Jeans length
will continually decrease, as will the Courant timestep, the largest timestep on which the 
hydrodynamical evolution can be followed while still maintaining numerical stability. Therefore,
once the gas reaches very high densities, the simulations can take only very small timesteps, 
and it is common practice in the numerical study of population III star formation to terminate
the simulations at this point. However, this practice means that the simulations can follow the 
evolution of multiple collapsing objects only if the collapses are very closely synchronized in
time. In reality, we know from the numerical study of local star formation that gravitational 
fragmentation is rarely so well synchronized. Typically, there is always some region with a
higher density, or a lower angular momentum, that collapses first, with other objects forming
only after a few local dynamical times. 
For example, the overall duration of star formation in nearby molecular
clouds is found to be comparable to the global crossing time of the cloud
\citep[e.g.][]{elm00,mk04} and exceeds the collapse timescale of individual 
stars by one to two orders of magnitude.
If only the initial collapse is simulated, the
formation of these other objects can be missed. To avoid this problem, it is common in numerical
studies of local star formation to replace gas which has collapsed beyond the limiting resolution
of the simulation with artifical sink particles. These particles possess the mass and linear
momentum of the gas that they replace, and continue to interact with the surrounding gas via
gravity. They are able to accrete additional infalling gas, provided that it is gravitationally bound 
to them and comes within a preset accretion radius. However, they no longer
feel the effects of 
hydrodynamical pressure gradients, and the subsequent evolution of the gas incorporated into 
them is not followed \citep{bbp95}. By replacing high density, unresolved gas with sink 
particles, it becomes possible to follow the process of gravitational fragmentation for many dynamical
times.

Despite their wide usage in the study of present-day star formation, sink particles have been
used in only a few studies of primordial star formation. \citet{bcl99,bcl02} used sink particles
in their study of Pop.~III star formation, creating them once the gas density exceeded 
$n_{\rm th} = 10^{8} \: {\rm cm^{-3}}$. They found that several massive clumps formed in
most of their simulations. The only case for which this was not true was for the smallest mass 
halo they simulated ($M_{\rm tot} = 2 \times 10^{5} \: {\rm M_{\odot}}$), in which pressure
forces would be expected to have the greatest effect. Taken at face value, these results 
suggest that the formation of several massive stars per minihalo could be a common outcome
of population III star formation. However, the initial conditions used in these simulations 
-- specifically, the adoption of a solid-body initial rotation profile -- have been criticized on 
the grounds that they are overly prone to fragmentation \citep{jkgm07}.  The large, 
rotationally-supported disks formed in the \citet{bcl99,bcl02} simulations are not seen 
in simulations that start from self-consistent cosmological initial conditions \citep[e.g.][]{abn02,
yoha06}, and so the fragmentation may also not occur.

A second study to utilize sink particles was that of \citet{bl04}. They adopted similar initial
conditions to \citet{bcl02}, but used a numerical technique called particle splitting 
\citep{kw02,bl03} to allow them to follow the evolution of the first dense clump up to a 
much higher density ($n_{\rm th} = 10^{12} \: {\rm cm^{-3}}$). They found no evidence for
sub-fragmentation of this dense clump on timescales $t \simless 10^{4} \: {\rm yr}$ after
the formation of the central sink particle. 
 
More recently, \citet{cgk08} used SPH with sink particles to simulate the collapse of
dense protostellar cores of various metallicities. Their simulations followed collapse from
an initial density of  $5 \times 10^{5} \: {\rm cm^{-3}}$ up to a  density of $n_{\rm th} \sim 
10^{17} \: {\rm cm^{-3}}$. 
The highest resolution simulation used 25 million SPH particles to represent $500 \: {\rm 
M_{\odot}}$, and so had a mass resolution of  $M_{\rm res} = 2 \times 10^{-3} \: {\rm M_{\odot}}$. 
The thermodynamic evolution of the gas was treated using a tabulated equation of state, 
based on the one-zone results of \citet{om05}. The initial rotational and turbulent energies
were chosen to be consistent with the results of previous studies of Pop.~III star formation,
such as \citet{abn02}. Although the primary focus of Clark \etal's study was an
examination of the effects of metal enrichment,  they also modelled a $Z = 0$ core for
comparison. They found that even in the primordial case, the core fragmented, forming of
the order of 20 sink particles. The mass function of these fragments was considerably flatter 
than the present day IMF, implying that most of the mass was concentrated in the few most 
massive fragments. Clark \etal\ also found that there was a delay of several local free-fall 
times between the formation of the first and second sink particles, and that at the time that 
the first sink particle formed, the radial profiles of mass density and specific angular 
momentum were similar to those seen in previous high-resolution simulations performed
without sink particles. Slices through the densest structure at the time that the first sink
particle forms also show little sign of the impending fragmentation (see e.g.\ Figure~1).
These results support the view that simulations without sink particles (or some comparable 
treatment of unresolved gas) run the risk of missing the formation of all but the first 
protostar.

\begin{figure}
\begin{center}
\unitlength1cm
\includegraphics[width=3in,height=3in]{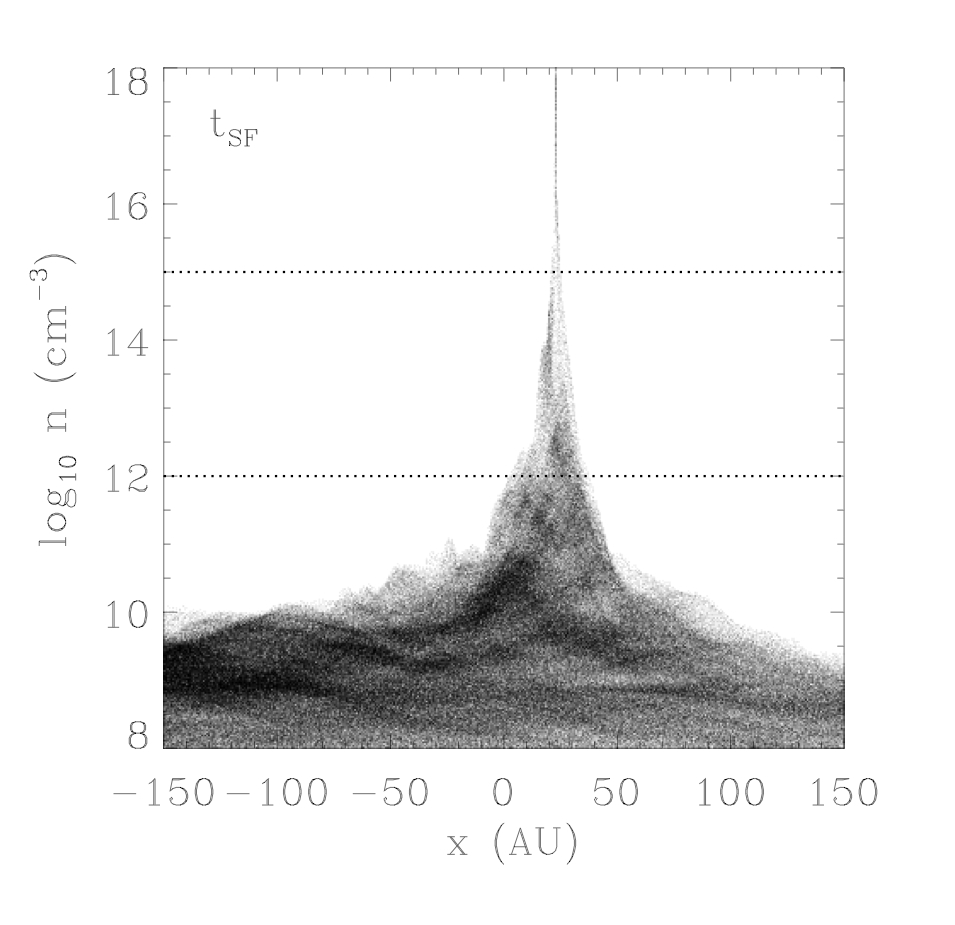}
\end{center}
\caption{The densities of the SPH particles in the \citet{cgk08} high resolution simulation, 
plotted as a function of their x-position at the moment that the first sink particle forms.
Only the particles contained within the central 300~AU of the simulation are shown. At this
point, there is little sign of the secondary fragmentation that will shortly occur.}
\end{figure}

Careful analysis of the \citet{om05} equation of state for
zero-metallicity gas shows roughly isothermal behavior in the density range
$10^{14}\,$cm$^{-3} \le n \le 10^{16}\,$cm$^{-3}$, i.e.\ just before the gas becomes
optically thick and begins to heat up adiabatically. Conservation of angular momentum
during this phase of the collapses leads to the build-up of a rotationally supported massive 
disk-like structure, which becomes gravitationally unstable and fragments.  
This is understandable, as isothermal disks are susceptible to gravitational instability 
\cite[][]{bodenheimer95} once they have accumulated sufficient mass. Further, 
\citet{goodwin04a, goodwin04b} show how even very low levels of turbulence can induce 
fragmentation. Since turbulence creates local anisotropies in the angular momentum on 
all scales, it can always provide some centrifugal support against gravitational collapse. 
This support can then provide a window in which fragmentation can occur. Moreover, the
density at which this occurs is significantly higher than the maximum density resolved in
the earlier \citet{bl04} simulation, and we would therefore not expect the fragmentation to
have been observed in that simulation.

However, the results of the \citet{cgk08} simulations come with several caveats attached.
The most important relates to the use of a tabulated equation of state to represent the
thermodynamic evolution. In this approach, one is essentially assuming that as the 
density of the gas changes, the gas temperature changes instantaneously to reflect the behaviour
prescribed by the equation of state. In reality, the gas temperature will adjust itself on a
timescale $t_{\rm cool}$, which in primordial gas is comparable to or greater than the dynamical 
timescale. This delay may damp out density fluctuations, thereby helping to suppress fragmentation.
A second concern is that the \citet{cgk08} simulations do not allow sink particles to merge,
and so may overestimate the number of fragments that survive in reality. However, there are
two reasons for believing that this is not a major concern. Firstly, the sink particle volume is 
considerably larger than the actual protostellar object in its center and so the cross section for 
two real protostars to collide is orders of magnitude smaller than the geometric cross section of 
the sink particles. Secondly, \citet{cb08} have demonstrated that the importance of collisions depends 
on the balance between shinkage of the cluster core by adiabiatic contraction and puffing via collisional 
relaxation. As a result of this balance, collisions will start to affect the mass function in the
\citet{cgk08} cluster only after $10^{3}$ to $10^{4}$ objects have been formed, further along in its 
evolution than has yet been simulated.

A final concern is that the Clark \etal\ simulations do not account for the effects of feedback
from the protostars that have already formed. Since fragmentation seems to occur on a timescale
much shorter than the protostellar Kelvin-Helmholtz timescale, feedback from the protostar is
probably unimportant. However, this assumption needs to be verified.

To summarize, there is some evidence, primarily from the work by \citet{bcl02} and \citet{cgk08},
that the number of population III stars that form in the earliest minihalos is higher than 
the single star that is commonly assumed. However, this conclusion remains controversial, and
the evidence is not yet convincing. 

\subsection{Is there a population III.2?}
\label{pop32}
The final issue that we discuss in this review is the question of whether there might be
more than one mode of population III star formation. As we have already discussed,
the first Pop.~III stars form in minihaloes in which $\mHt$ dominates the cooling,
with HD generally playing only a minor role. However, work by a number of authors has shown
that if molecular hydrogen formation can be more efficiently catalyzed by a higher electron
abundance, then HD cooling of the gas can become efficient, and can allow the gas
temperature to reach values close to the floor set by the CMB \citep[see e.g.][]{nu02,no05,jb06,yokh07}.  
Efficient HD cooling lowers the characteristic gravitational fragmentation mass scale, and 
may also reduce the accretion rate of gas onto the fragment or fragments that form
\citep{yokh07,mb08}.
It is therefore reasonable to assume that if HD cooling becomes dominant, lower mass
stars will be formed than in the case in which $\mHt$ dominates throughout. If this is true,
then it suggests that there may be two distinct sub-populations of stars within population
III: a first generation of stars forming from undisturbed primordial gas (termed population III.1
by \citealt{tm08}) and a subsequent generation forming from gas that remains metal-free
but that has an elevated fractional ionization, owing to either the infall of
the gas into the deeper potential wells of the first galaxies, or to the
effects of feedback from the first generation of stars.
\citet{tm08} term this second generation `population III.2'. 

There are various different scenarios that may lead to the formation of population III.2 
stars. The simplest scenario involves primordial star formation in the first galaxies
\citep{gb06}.
If the virial temperature of such an object exceeds $10^{4} \: {\rm K}$, then most infalling
gas will be shock-heated to temperatures high enough for it to become ionized. As this gas
subsequently cools and recombines, it will form $\mHt$ at an accelerated rate, owing to
the enhanced fractional ionization of the gas \citep{sk87}. If enough $\mHt$ is formed to
cool the gas to roughly 150~K, then HD cooling will take over, driving the temperature down
towards the CMB floor. Recent simulations of the formation of the first galaxies
by \citet{gjkb08} show this mechanism in operation (see Fig.~2). However, this scenario can only produce
population III.2 stars if the gas forming these galaxies has remained 
metal-free. \citet{gjkb08} demonstrate that these galaxies typically have of order 10 or
more progenitors that have undergone population III.1 star formation, and so it is likely
that most of their gas will already have been contaminated with metals. Nevertheless, 
population III.2 stars may still be able to form in these objects if metal mixing is 
inefficient, or if some fraction of the gas that they accrete has remained pristine.

\begin{figure}
\begin{center}
\includegraphics[width=13cm]{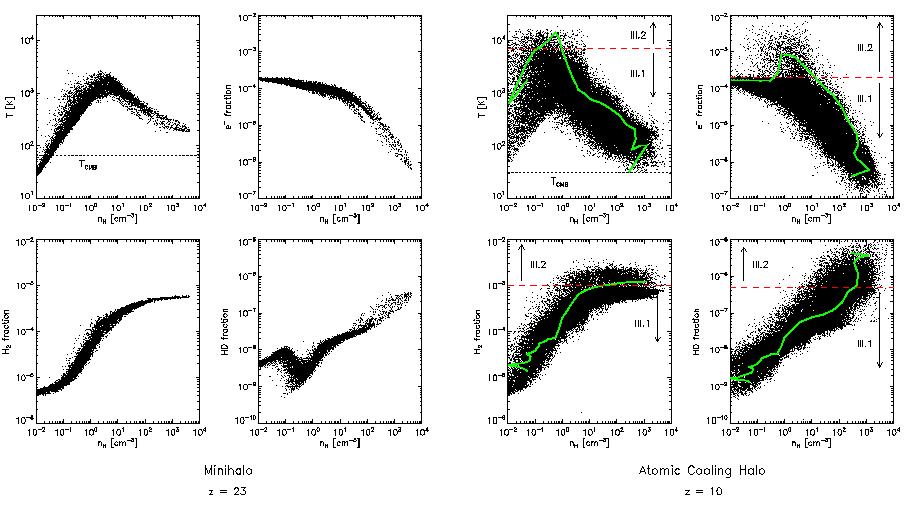}
\caption{The phase-space distribution of gas inside a minihalo (left-hand panel) and an atomic cooling halo (right-hand panel). We show the temperature, electron fraction, HD fraction and H$_{2}$ fraction as a function of hydrogen number density, clockwise from top left to bottom left. {\it Left-hand panel:} In the minihalo case, adiabatic collapse drives the temperature to $> 10^{3}~\rm{K}$ and the density to $n_{\rm{H}}> 1~\rm{cm}^{-3}$, where molecule formation sets in and allows the gas to cool to $\simeq 200~\rm{K}$. At this point, the central clump becomes Jeans-unstable and ultimately forms a Pop~III.1 star. {\it Right-hand panel:} In the galaxy, a second cooling channel has emerged due to an elevated electron fraction at the virial shock, which in turn enhances molecule formation and allows the gas to cool to the temperature of the CMB. The dashed red lines and arrows approximately delineate the resulting Pop~III.1 and Pop~III.2 channels, while the solid green lines denote the path of a representative fluid element that follows the Pop~III.2 channel.}
\end{center}
\end{figure}

The Pop~III.2 star formation mode may also be triggered by the explosions of the first supernovae (SN), 
as the shocks from these explosions can heat and ionize primordial gas \citep[e.g.][]{mbh03,sfs04,mach05,jb06}.  In a three-dimensional simulation 
of the explosion of a Pop~III pair-instability SN, \citet{gjbk07} found that the SN shock-compression
of the gas in minihalos can speed its collapse.  However, in this simulation the neighboring minihalo was 
not strongly shocked and so did not yield any evidence for the formation of Pop~III.2 stars, although it 
is noted that minihalos within perhaps $<$ 500 pc of such a SN could become strongly ionized.  Such 
minihalos would be the likely sites of the formation of Pop~III.2 stars which are formed in the wake of 
the first SN.  Simulations of SN feedback on close-by minihalos which account for the mixing of the 
metals injected by the supernova with the primordial gas are an important next
step \citep[see][]{cr07}.

A further location in which population III.2 stars may be formed is in the relic 
H$\,${\sc ii} regions left behind after the death of the first massive Pop.~III stars.
Again, the gas starts in a hot and highly ionized state, and forms significant quantities
of $\mHt$ as it cools and recombines \citep{rgs02,jgb07,yoh07}. However, the H$\,${\sc ii} regions 
produced by the first massive stars are generally capable of expelling most of the gas
from the galaxies containing the stars, and expanding to fill large volumes in the 
low-density intergalactic medium. Therefore, most of the $\mHt$ that forms within the
relic H$\,${\sc ii} regions resides at very low densities and is initially unavailable
for star formation. 

Finally, Pop.~III.2 stars may also be formed in primordial gas irradiated with a 
sufficiently strong flux of cosmic rays. \citet{sb07} studied the effects of cosmic rays
on population III star formation and showed that the enhanced fractional ionization that
they create would lead to enhanced $\mHt$ and HD production and the cooling of the gas
to $T \sim T_{\rm CMB}$ for cosmic ray ionization rates greater than $\zeta \sim 10^{-19} 
\: {\rm s^{-1}}$. \citet{jce07} reached similar conclusions in a separate study.

As all three scenarios rely on the gas forming more $\mHt$ than in the standard Pop.~III.1 
formation scenario, additional physical effects that reduce the amount of $\mHt$ formed 
will lessen the likelihood that any Pop.~III.2 stars form. One such effect is radiative
cooling by $\mHt$-$\Hp$ and $\mHt$-$\me$ collisions. \citet{ga08} include these processes
in their models of primordial gas cooling and show that, somewhat counter-intuitively, 
the increased cooling that they provide at early times (while the fractional ionization
is high) leads to less $\mHt$ being formed, and hence to less cooling at late times. 
This surprising result is a consequence of the different temperature dependences of the
rate coefficients for $\Hm$ formation and for $\Hp$ recombination. Decreasing the 
temperature decreases the rate at which $\Hm$ forms, and hence decreases the $\mHt$ 
formation rate. However, it also increases the $\Hp$ recombination rate, and so reduces
the time available for $\mHt$ formation before the necessary electrons are lost from
the gas. Therefore, the faster the gas cools at early times, while the fractional 
ionization remains large, the less $\mHt$ it will ultimately form. \citet{ga08} show
that in their one-zone calculations, this effect does not prevent the gas from cooling
below 100~K; however, its effects in more realistic situations have not yet been
investigated.

Another obvious candidate for suppressing Pop~III.2 star formation is the far ultraviolet 
background built up by the first stars \citep{har00,mba01,rgs02,ahn08}. This dissociates 
$\mHt$ and HD, and so acts to suppress cooling. \citet{yoh07} estimate that a far-UV flux of only 
$3 \times 10^{-22} \: {\rm erg} \: {\rm s^{-1}} \: {\rm cm^{-2}} \: {\rm Hz^{-1}} \:
{\rm sr^{-1}}$ is required in order to prevent the gas from cooling below 200~K, thereby
completely suppressing the formation of Pop.~III.2 stars. However, this estimate is based
on simple one-zone calculations that are unlikely to properly capture the dynamics of the
gas, and so may be misleading.  
Furthermore, in galaxies in which the virial temperature exceeds $\sim$ 10$^4$ K
the column density of H$_2$ may become high enough to shield the central regions from 
the far ultraviolet background radiation, thereby allowing HD molecules to survive and 
to be an important coolant (see Johnson et al. 2008).  An investigation 
of the effects of the ultraviolet background on HD cooling and the formation of Pop.~III.2 
using a fully three-dimensional approach would be very valuable.

\section{Outlook}
Of the four open questions discussed in this review, those involving 
uncertainties in the chemistry and in the cooling of metal-free gas
seem the easiest to address. Existing work has gone a long way towards
establishing the effects of the chemical uncertainties, and scientists
from the atomic and molecular physics communities are rising to the
challenge that these uncertainties present. As far as the cooling is
concerned, we now have a basic understanding of which processes are 
important and which are unimportant over a very wide range of scales
during protostellar collapse. While there may still be occasional
surprises, such as the importance of $\mHt$-$\Hp$ and $\mHt$-$\me$
collisions in gas with only a slightly elevated fractional ionization,
we think it unlikely that any of these will fundamentally change our
picture of population III star formation.

The question of how many population III stars form in each minihalo
is much further from being settled. There is now reason to believe
that the conventional wisdom that only one massive star forms per
minihalo may be incorrect. On the other hand, it may be that it is
the simulations that are incorrect; they may be giving us a misleading
view of what happens owing to the approximations that they make.
To settle this question, further numerical study is required, using
methods that are capable of following the hydrodynamical evolution
of the gas beyond the point at which the first protostar forms, but
that do not make as many approximations as in the \citet{cgk08} 
study. 

Finally, the question of whether Pop.~III.2 exists as a distinct
sub-population within population III also presents continuing 
difficulties. Some uncertainties, such as the impact of the revised
treatment of $\mHt$ cooling presented by \citet{ga08}, will be
easy to address with the next generation of numerical simulations,
and so should be resolved within the next year or so. However,
other issues, such as the impact of the extragalactic far ultraviolet
background, involve physics that is difficult to simulate
accurately, and it will take far longer before we fully understand
its effects. (As an example, consider that after more than ten years
of study, there is still not complete agreement regarding the
degree to which the ultraviolet background regulates $\mHt$ cooling
in minihalos; c.f.\ \citealt{hrl97,har00,rgs02,wa07,on08}). 
Furthermore, even after these issues are addressed, we will still
not be able to claim with confidence that Pop.~III.1 and Pop.~III.2
differ until we have a better understanding of the processes
regulating accretion onto population III stars. Ultimately, this
may be a question that is answered as much by stellar archaeology 
as by theoretical study.

\acknowledgements
The authors would like to thank the organisers of IAU Symposium 255
for organising a very stimulating and enjoyable meeting. RSK acknowledges 
partial support from the Emmy Noether grant KL 1358/1. RSK, TG, and 
PCC also acknowledge support from the DFG SFB 439 `Galaxies in the
Early Universe'. TG would also like to thank the Heidelberg Graduate School 
of Fundamental Physics (HGSFP) for financial support. The HGSFP is funded by
the excellence initiative of the German government (grant number GSC 129/1).
VB acknowledges support from NSF grant AST-0708795.

\end{document}